\documentclass[12pt,epsf]{article}
\usepackage{graphicx}

\begin{document}

\sloppy
\begin{flushright}{SIT-HEP/TM-21}
\end{flushright}
\vskip 1.5 truecm
\centerline{\large{\bf Formation of cosmological brane defects}}
\vskip .75 truecm
\centerline{\bf Tomohiro Matsuda
\footnote{matsuda@sit.ac.jp}}
\vskip .4 truecm
\centerline {\it Laboratory of Physics, Saitama Institute of
 Technology,}
\centerline {\it Fusaiji, Okabe-machi, Saitama 369-0293, 
Japan}
\vskip 1. truecm
\makeatletter
\@addtoreset{equation}{section}
\def\theequation{\thesection.\arabic{equation}}
\makeatother
\vskip 1. truecm

\begin{abstract}
\hspace*{\parindent}
We study cosmological formation of D-term strings, axionic strings,
 domain walls 
and Q-balls in braneworld models of the Hanany-Witten type.
For the D-term strings, we show that the strings are the daughter branes
extended between mother branes.
We show that the domain walls can be produced by conventional
 cosmological phase 
transitions.
In this case, the formation of the domain walls is induced by the
continuous deformation of the branes, which means that they are not
 created as daughter branes.
First we consider classical configurations of the axionic strings
and the domain walls, then we investigate the quantum effect of the brane
 dynamics.
We also study brane Q-balls and show how they can be distinguished
from conventional Q-balls. 
\end{abstract}

\newpage
\section{Introduction}
Although there were great successes in quantum field theory, we
still have no consistent scenario in which the quantum gravity is included.
The most promising scenario in this direction will be string
theory where the consistency is ensured by the requirement of
additional dimensions and supersymmetry.
The idea of large extra dimension\cite{Extra_1} may solve or weaken
the hierarchy problem. 
In this case, denoting the volume of the $n$-dimensional compact space
by $V_n$, 
the observed Planck mass is obtained by the relation $M_p^2=M^{n+2}_{*}V_n$,
where $M_{*}$ denotes the fundamental scale of gravity.
The standard model fields are expected to be localized
on a wall-like structure, and the graviton propagates in the bulk.
The natural embedding of this picture in the string theory context 
is realized by a brane construction.
Inflation with such a low
fundamental scale is still an interesting topic\cite{low_inflation,
matsuda_nontach, matsuda_defectinfla}.
Other cosmological issue such as baryogenesis with low fundamental
scale is discussed in ref.\cite{low_baryo, Defect-baryo-largeextra,
Defect-baryo-4D}, where cosmological defects play important roles.
Constructing models for the particle cosmology
where non-static brane configurations (such as brane defects and
Q-balls\cite{BraneQball}) are very important.
We are expecting that future cosmological observations will reveal
the evolution of the Universe, which might also reveal the physics beyond
the standard model.
To know what kind of brane defects are allowed in the evolution of the 
Universe, we need to understand how they are formed (and disappeared) in the
history of the Universe.  
In the original scenario for brane inflation\cite{brane-inflation0},
the inflationary expansion is driven by the potential between branes
and anti-branes evolving in the bulk space of the compactified
dimensions.
The end of inflation is induced by the brane collision 
where the brane annihilation proceeds through tachyon
condensation\cite{tachyon0}. 
During brane inflation, tachyon is trapped in the false vacuum.
Then the tachyon starts to condensate after inflation, which may result
in the formation of the daughter branes. 

The production of cosmological defects after brane inflation is discussed
in ref.\cite{Brane-defects, angled-defect}, where it is concluded that
cosmic strings are copiously produced but the domain
walls are negligible.
In ref.\cite{Majumdar_Davis}, however, it is discussed that all kinds 
of defects can be produced and the conventional problems of cosmic
domain walls and monopoles should arise.
Later in ref.\cite{D-brane-strings, Halyo, BDKP-FI}, the brane
production is reexamined 
and the conclusion was different from \cite{Brane-defects,angled-defect} and
\cite{Majumdar_Davis}. 
In ref.\cite{Brane-defects, angled-defect}, it is discussed that the
effect of compactification 
is significant for the defect formation due to tachyon condensation.
It must be useful to make a brief review of the previous arguments about
the cosmological formation of brane defects.
Their argument is that since the compactification radius is small
compared to the horizon size during inflation, any variation of a 
field in the compactified direction is suppressed.
Then the daughter brane wraps the 
same compactified dimensions as the mother brane.
As a result, the codimensions of the daughter branes lie within the
uncompactified space.
Since the number of the codimension must be even, the
defect is inevitably a cosmic string.
Moreover, in ref.\cite{D-brane-strings}, it is pointed that the analysis
does not fully account for the effect of compactification, since the
directions transverse to the mother brane is not considered.
The effect of the RR fields extended to the compactified dimensions is
discussed in ref.\cite{D-brane-strings}.
The result is that the creation of the gradients of the RR fields in the
bulk of the compactified space is
costly in energy, so that the creation of the daughter brane is
suppressed if it does not fill all the compactified dimensions.
In this case, it was concluded that the production of cosmic strings
requires efficient mechanisms, 
and monopoles and domain walls are not produced after brane
inflation.\footnote{Another kind of defects, which are parameterized
by the positions of the branes, were constructed in
ref.\cite{Alice-string} and later in ref.\cite{incidental}.
In ref.\cite{Alice-string}, brane is
replaced by a domain wall that is embedded in the higher-dimensional spacetime,
so that one can see what happens in the core. 
Then the position of a brane in the 
fifth dimension is used to parameterize the cosmic string in the
effective four-dimensional spacetime. 
The brane is shown to be smeared in the core, so that it resolves the
anticipated singularity.  
Then in ref.\cite{incidental}, the {\bf relative} positions between
branes are used. 
In ref.\cite{incidental}, these defects are called incidental brane defects.
In these field-theoretical constructions, branes are replaced by domain
walls or vortices embedded in the higher-dimensional spacetime.}

In this paper, however, we show explicit examples where the production
of cosmic strings is realized by the formation of daughter branes
that are extended between splitting mother branes. 
It should be noted that we are not considering a counter example of the
mechanism of tachyon condensation. 
The problem of the RR field is avoided, since the length of the extended
daughter brane vanishes when it is formed.
The tension of the fully extended daughter brane matches to the tension
of the D-term string in the effective Lagrangian.\footnote{In our next
paper\cite{matsuda_angleddefect}, we consider another type of angled
brane inflation and solve the problem of the $\theta$-dependence of the
string tension.}
Moreover, we also show that other cosmological defects, such as domain
walls and Q-balls, can be produced after brane inflation.
In our model, it is natural to think that the domain walls and the Q-balls are
not produced by the brane creation.\footnote{Of course it is not impossible to
think that such domain walls are the daughter branes being extended
between vacuum branes.
In our case, however, the cosmological evolution of the brane
configuration suggests that they are not produced by the creation of 
daughter branes.
Cosmological formation of domain walls and monopoles, which is induced
by the creation 
of daughter branes, will be discussed in ref.\cite{matsuda_future}.
In this case, the production of the branes extended between branes is
crucial.}
Domain walls are corresponding to the spatial deformations of the vacuum
branes, which can be formed by the thermal effect\cite{thermal-brane}
or brane oscillation after inflation.
Thermal effect can induce attractive forces between branes.
Then the observer in the four-dimensional spacetime sees the restoration
of the corresponding symmetry.
The spontaneous breaking of the symmetry triggers the formation of
the cosmological defects, which is parameterized by the relative position
between branes.

In Section 2, we begin with a short review of a
model\cite{HHK-braneinflation} for brane inflation 
due to the Hanany-Witten\cite{Hanany-Witten} type brane
dynamics. 
We show how the extended branes are produced by the brane dynamics
after brane inflation.
Although our result seems to be contradicting to the previous
arguments about daughter brane production, we stress that we are not
considering a 
counter example of the mechanism of tachyon condensation.
We think it is not difficult
to understand how one can avoid the serious criteria given in
ref.\cite{Brane-defects, D-brane-strings}.
The extended branes are formed after brane inflation.
The correspondence between the extended brane and the D-term
string in the effective action is examined.  
Then, the formation of axionic strings and domain walls is discussed in
Section 3. 
Unlike the D-term strings, these defects
are formed by the spatial deformations of the vacuum branes.
The domain wall that we are considering in this paper is different from 
the usual BPS domain walls\cite{Witten_Wall} in SQCD and MQCD. 
In the effective action, we add a soft mass for the adjoint scalar field
in $N=2$ SYM, 
which introduces a shallow potential on the coulomb branch.
In the brane counterpart, we are considering $D4$-branes separated by a
weak repulsive force between them.
Thus our defect configurations are not stable in the supersymmetric
limit. 
Our discussions in this paper compensate the analysis in 
ref.\cite{Alice-string} and \cite{incidental}, in which defects were
constructed in the classical brane configurations.
It should be noted that the conventional BPS domain walls in MQCD are not
suitable for our argument, because they cannot exist in the classical
limit.
We also consider brane Q-ball\cite{BraneQball}, which
is the configuration of branes in motion.
Our conclusions and discussions are given in Section 4. 

\section{D-term string produced after brane Inflation}
\hspace*{\parindent}
In this section, we discuss the formation of D-term strings.
We show explicitly why it is possible to produce daughter 
branes that are extended between mother branes.

\subsection{4D effective Lagrangian and D-term strings}
The form of the 4D hybrid potential for the P-term inflation model
is\cite{HHK-braneinflation}
\begin{equation}
\label{D-term-potential}
V=\frac{g^2}{2}\left[
\left(|\Phi_1|^2 + |\Phi_2|^2\right)|\Phi_3|^2
+|\Phi_1|^2|\Phi_2|^2
+\frac{1}{4}\left(|\Phi_1|^2 - |\Phi_2|^2 + \frac{2\xi}{g}\right)^2
\right]
\end{equation}
Here the two complex scalar fields $\Phi_1$ and $\Phi_2$ form a
quaternion of the hypermultiplet, which is charged under the $U(1)_X$
group.
The complex scalar field $\Phi_3$ appears from the $N=2$ vector
multiplet.
$\xi$ is the Fayet-Iliopoulos (FI) term for the $U(1)_X$ symmetry.
When the vacuum expectation value (vev.) of $|\Phi_3|$ is large, the vev of
the hyper multiplets vanish and the $|\Phi_3|$ direction becomes flat.
In this case, $\Phi_3$ is the inflaton of hybrid inflation, whose
potential is lifted by the logarithmic corrections.
The trigger field is $\Phi_2$, which roll toward
true vacuum and reheat the Universe at the end of inflation.

The D-term strings are formed at the end period of inflation.
Note that $\Phi_3$ takes large value during inflation, which stabilizes
the potential for $\Phi_1$ and $\Phi_2$ to make these fields to stay at
their origin.
At the end of inflation, when $\Phi_3$ becomes smaller than
a critical value, the potential is destabilized and the trigger field
starts to roll down the potential. 
The D-term string is formed at the end of inflation, by the spontaneous
breakdown of the $U(1)_X$ symmetry.
It is easy to calculate the scaling property of the tension of the
D-term strings from the 
potential in eq.(\ref{D-term-potential}),
which becomes\cite{D-brane-strings, Vilenkin-book}
\begin{equation}
\label{D-strin-tension}
T_{D-string} = \frac{2\pi \xi}{g}.
\end{equation}
The BPS conditions for the D-term strings are examined in
\cite{D-termstring}, in which the D-term strings are shown to 
satisfy the BPS conditions in supergravity.

\subsection{D2 string in NS5-D4/D6-NS5 model}
\begin{figure}[ht]
 \begin{center}
\begin{picture}(300,200)(0,0)
\resizebox{10cm}{!}{\includegraphics{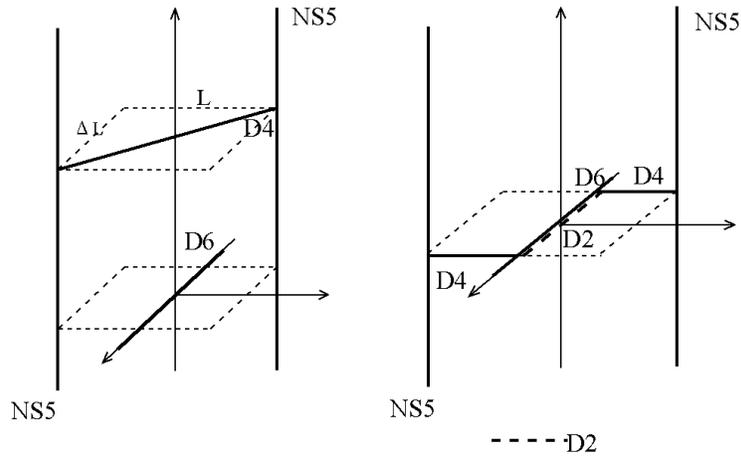}} 
\end{picture}
\caption{A schematic representation of brane
  inflation in ref.\cite{HHK-braneinflation}. 
The splitting of the D4-brane is due to the rotation between the
  D4-brane and the D6-brane,
which induces the constant Fayet-Iliopoulos term in the effective
  Lagrangian. 
Dotted line on the D6-brane denotes 
the D2-brane, which corresponds to the D-term string in the effective
  Lagrangian. }  
\label{D2_in_HHK}
 \end{center}
\end{figure}

The D-term strings that are produced after P-term inflation 
have the tension of the form (\ref{D-strin-tension}), and satisfy the BPS
conditions. 
The possible brane counterpart is
the D2-brane, which is extended between the split
D4-branes.\footnote{See fig.\ref{D2_in_HHK}.} 
From the viewpoint of the effective Lagrangian, the production of the
D-term strings after inflation is not suppressed.
However, as is noted in ref.\cite{Brane-defects,angled-defect,
D-brane-strings}, the 
effect of compactification seems to be significant for the defect
formation due to tachyon condensation. 
Since the compactification radius should be small compared to the
horizon size during inflation, any {\bf cosmological} variation of a
field in the compactified direction must be suppressed.
If there is {\bf no other mechanism} that induces variation of the field, 
the daughter brane must wrap the same compactified dimensions as the
mother brane. 
As a result, the codimensions of the daughter branes seem to lie within the
uncompactified space.
Moreover, in ref.\cite{D-brane-strings}, it is pointed that the analysis
does not fully account for the effect of compactification, since the
directions transverse to the mother brane is not considered.
Then in ref.\cite{D-brane-strings}, the effect of the RR fields that are
extended to the compactified dimensions is discussed.
Their conclusion is that the creation of the gradients of the RR fields
in the bulk of the compactified space is
costly in energy, so that the creation of the daughter brane is
suppressed even if they wrap the same compactified dimensions as the
mother brane.

\begin{figure}[ht]
 \begin{center}
\begin{picture}(350,200)(0,0)
\resizebox{13cm}{!}{\includegraphics{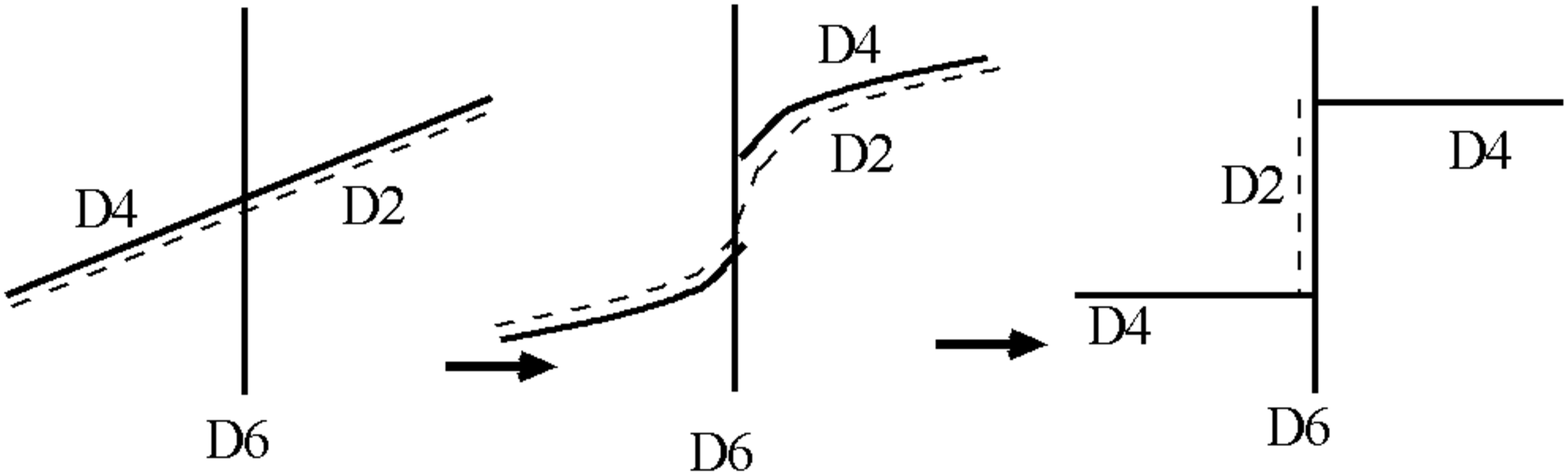}} 
\end{picture}
\caption{Brane inflation ends when the $D4$ brane splits on the $D6$
  brane. The dotted line in the left picture is the seed for the $D2$
  brane. To be more precise, a careful treatment of the effective
  action\cite{Localize_tachyon} shows that the eigenfunction of the
  tachyonic mode is localized at the intersection. Since the mechanism
  of this localization is {\bf different} from the Kibble mechanism, the
  ``seed'' for the $D2$ brane can be localized at the intersection. 
  Then the $D2$ brane is pulled out from the mother brane when
  the mother $D4$ brane splits on the $D6$ brane.}   
\label{D2_in_HHK2}
 \end{center}
\end{figure}

Is it really impossible to produce extended branes by the daughter
brane production?
Let us see more detail.
If the initial perturbation of the tachyon condensation produces the
seed for the $D2$ branes on the world volume of the mother brane, it
might wrap the same 
compactification space as the mother brane.
The dotted line in the left picture in fig.\ref{D2_in_HHK2} denotes the
``seed'' for the $D2$ brane.
To be more precise, a careful treatment of the effective
action\cite{Localize_tachyon} shows that the eigenfunction of the tachyonic 
mode is localized at the intersection.
Since the mechanism of this localization is {\bf different} from the Kibble
mechanism, the ``seed'' for the $D2$ brane can be localized at the
intersection. 
From the string perspective, it seems obvious that the
brane creation is due to the local dynamics of the open strings at the
D4-D6-D4 intersection of the splitting D4 branes.
As the recombination proceeds, the $D2$ brane is
pulled out from the mother $D4$ brane, and finally becomes extended
between split mother branes.
In this case, the problem of the RR field is avoided since the length
of the extended 
$D2$ brane vanishes at the time when it is pulled out from the mother
brane.
Of course, it costs energy to pull $D2$ branes out from the $D4$ branes,
however in this case the cost is paid by the repulsive force between the
splitting $D4$ branes.
Our conclusion is consistent with the analysis of the effective action,
where the production of the D-term strings is not suppressed.

Let us examine if the tension of the fully extended brane matches to
the result obtained from the effective Lagrangian.
The brane splitting in the brane dynamics corresponds to the spontaneous
breaking of the $U(1)_X$ symmetry in the 4D effective Lagrangian.
We take the distance $\Delta L$ as in fig.\ref{D2_in_HHK}.
Then the tension of the string that corresponds to the extended D2-brane
becomes\cite{Brane-book} 
\begin{equation}
\label{D2-tension}
T_{D2}=\frac{\Delta L}{g_s(2\pi)^2 \alpha^{'\frac{3}{2}}}.
\end{equation}
Here $g_s$ and $l_s$ are the string coupling and the string
length.
We have defined that $\alpha' = l_s^2$.
Following ref.\cite{HHK-braneinflation}, the gauge coupling constant in
the effective four-dimensional Lagrangian 
is given by the formula,
\begin{equation}
\label{4d-coupling}
g^2=(2\pi)^2 g_s \frac{l_s}{L},
\end{equation}
where $L$ is given in fig.\ref{D2_in_HHK}.
As a result, the tension of the string in the effective action, which
corresponds to the D2-brane, is given by 
\begin{equation}
T_{D2}=\frac{\Delta L}{ \alpha' g^2 L}.
\end{equation}
The above result matches to the tension of the D-term string, since the
Fayet-Iliopoulos term in the effective Lagrangian is given
by\cite{HHK-braneinflation} 
\begin{equation}
\xi = \frac{\Delta L /L}{g 2\pi \alpha'},
\end{equation}
where $\Delta L \ll L$ is assumed.
Thus we can confirm that the tension of the D-term string matches to the
extended D2-brane, $T_{D2} = 2\pi \xi/g = T_{D-string}$.

\section{Domain walls, axionic strings and brane Q-balls from brane dynamics}
The defect formation in the brane world
does not always correspond to the production of the 
lower-dimensional branes.
For example, the position of a brane in the
fifth dimension can be used to generate cosmic
strings\cite{Alice-string}.
Later in ref.\cite{incidental}, it is shown that the relative position of
the branes in the higher-dimensional bulk space can fluctuate to form
such brane defects.
Explicit examples for domain walls, strings and monopoles
are shown in ref.\cite{incidental}.
These defects are formed by the deformations of the vacuum branes.
In ref.\cite{Alice-string}, the branes are
replaced by the domain walls that are embedded in the higher-dimensional
spacetime, so that one can see what happens in the core. 
Then it is shown that the branes are smeared in the core, so
that the 
anticipated singularity is resolved.  
Therefore, the field-theoretical construction is useful
to understand the 
inner structure of these defects.
However, if one considers the field-theoretical construction,
it is quite difficult to include the quantum effect of the brane
dynamics.
Therefore, it is important to formulate these types of brane
defects in the 
Hanany-Witten type brane dynamics, so that we can investigate the quantum
effect of the brane dynamics.
We think it is easy to understand that these defects are produced during
the usual 
cosmological evolutions of the Universe.
Although it is not impossible to regard that the domain walls are the
daughter branes being extended between mother branes,
from the cosmological viewpoint these domain walls are formed by the spatial
deformations of the vacuum 
branes during the usual cosmological evolution of the Universe.

\begin{figure}[ht]
 \begin{center}
\begin{picture}(300,200)(0,0)
\resizebox{10cm}{!}{\includegraphics{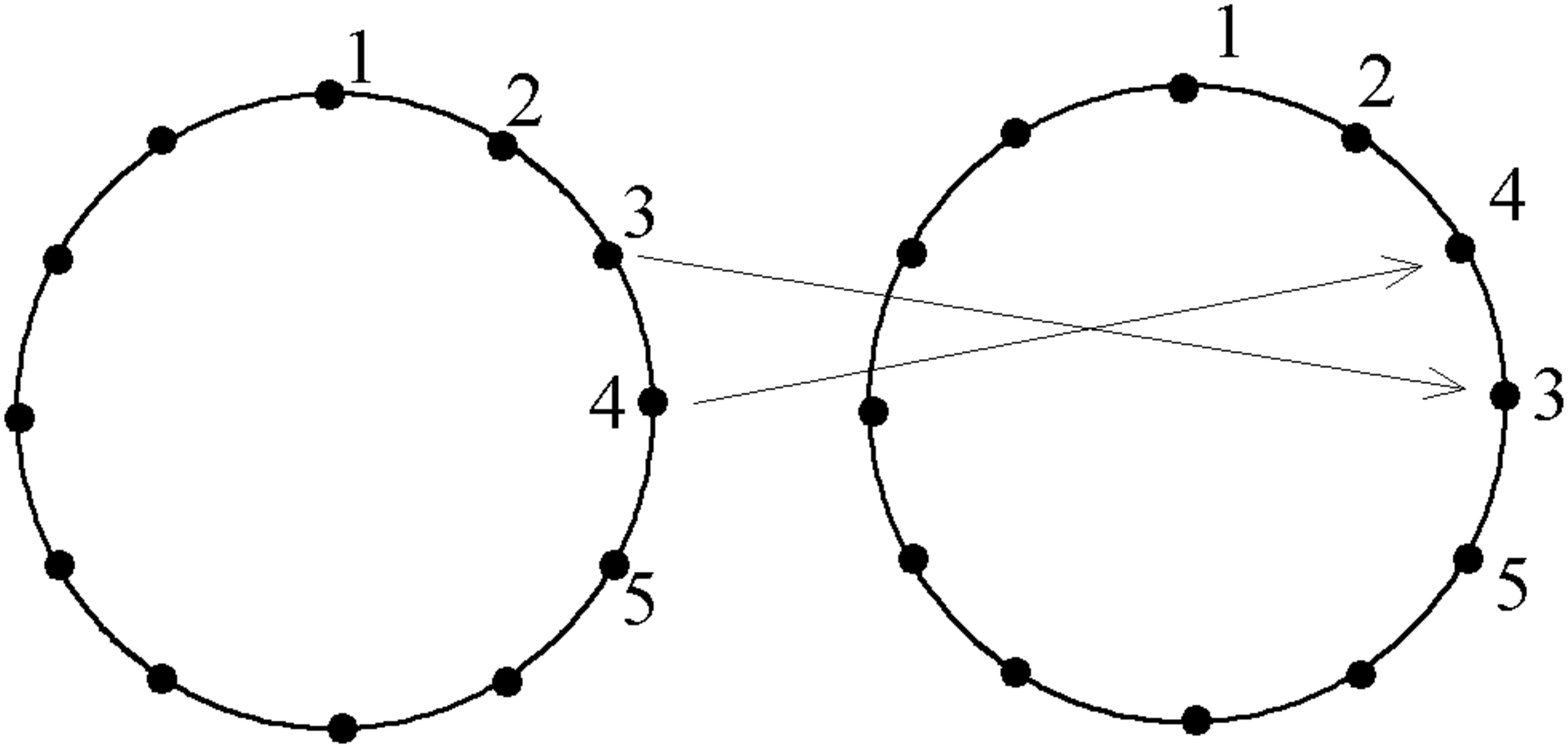}} 
\end{picture}
\caption{Permutation between the two branes ``3'' and ``4''.
If the two vacua are physically distinct, the domain wall, which
  interpolates between them, is produced.}  
\label{Permutation_Wall}
 \end{center}
\end{figure}

\begin{figure}[ht]
 \begin{center}
\begin{picture}(200,200)(0,0)
\resizebox{10cm}{!}{\includegraphics{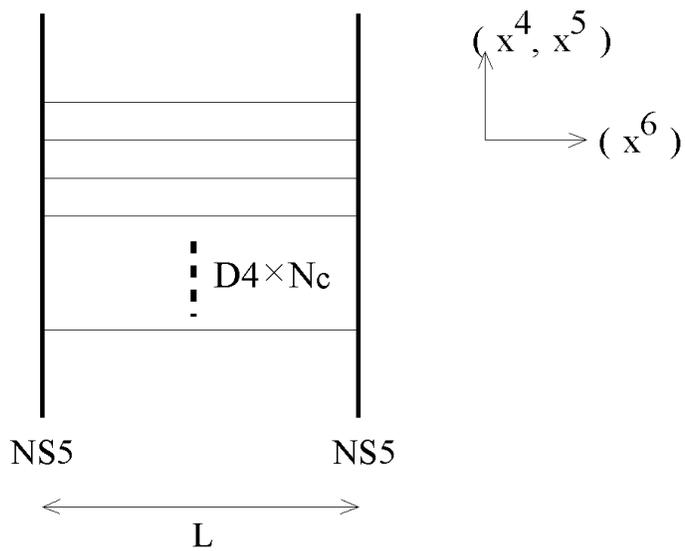}} 
\end{picture}
\caption{The brane configuration in Section 3. 
The parallel $N_c$ D4-branes are stretching between the two NS5-branes.
The endpoint of a D4-brane on the NS5-brane is on the $(x^4, x^5)$-plane.
In more realistic models, these configurations are assumed to be
  embedded in the compactified space.}  
\label{MQCD_NS5-D4}
 \end{center}
\end{figure}

First, we briefly describe our basic ideas for the incidental brane
defects in the classical brane configuration.
We consider parallel branes of codimension 2, and assume that
there is a weak repulsive force between them.
We also assume that the potential for the distance between branes is
stabilized at a distance.\footnote{The origin of the repulsive force is
 a weak supersymmetry breaking.
The stabilization at a distance is possible
if there are higher dimensional terms suppressed by the cut-off scale.}
Then we can consider a configuration in which the branes are located on a
circle, as is depicted in fig.\ref{basic-brane}.
Here we start from the classical arguments.
Seeing fig.\ref{basic-brane}, it seems natural to consider 
the following configurations.
\begin{itemize}
\item ``Domain wall that interpolates between the two degenerated vacua.''\\
      A domain wall is produced by the simple permutation between
      two branes. 
      The permutation is shown in fig.\ref{Permutation_Wall}.
\item ``Strings being formed by the rotation of their relative
      positions.''\\
      The configuration is shown in
      fig.\ref{Axionstring}. 
\item ``Brane Q-balls''\\
      In this case, unlike the above setups for axionic strings and
      domain walls, 
      the true vacuum is needed to be placed at the
      origin.\footnote{One can expect that the
      flat potential is lifted by a soft mass. One can also consider 
      a small $\mu$-term for the adjoint field. }
      The brane Q-ball is the configuration of the branes in motion, where
      the branes are rotating around each other\footnote{Although the basic
      idea for the brane Q-ball is not different from the previous
      discussions in ref.\cite{BraneQball}, the situation in
      this paper is more realistic than the simple brane anti-brane pair
      that is discussed in
      ref.\cite{BraneQball}}. 
\end{itemize}
In ref.\cite{incidental}, from the field-theoretical construction
(i.e., the branes are constructed as the embedded defects in the
higher-dimensional bulk), we have constructed
several types of incidental brane defects.
The brane Q-ball was discussed in ref.\cite{BraneQball}, in which the
simple brane anti-brane system was considered.
As we have discussed above, the field-theoretical construction
is convenient to see the structure in the defect core.
In ref.\cite{Alice-string}, the field-theoretical construction is used to show
how the expected singularity in the core is resolved by the smearing brane.
On the other hand, in the field-theoretical construction, it is
difficult to examine the quantum effect, which is induced by the
brane dynamics.  
Therefore, it is important to investigate the quantum effect of the brane
dynamics for the configurations of these brane
defects.
In this section, we try to construct the above-mentioned defects
in the setups of the Hanany-Witten type, where the quantum effect of
the brane dynamics is shown to play important roles for cosmological
strings and domain walls. 
Our arguments compensate the previous
results in ref.\cite{BraneQball, Alice-string, incidental}.
Our classical setup of the Hanany-Witten model is depicted in
fig.\ref{MQCD_NS5-D4}. 

Let us consider a global $U(1)$ symmetry, which is depicted in
fig.\ref{Axionstring}.
The $U(1)$ symmetry is the classical symmetry of the effective
four-dimensional action, but will be broken by the quantum effect.
Therefore, if a global string is constructed by the $U(1)$ symmetry, it 
should turn out to be a junction (or boundary) of the domain walls.
However, as far as one is considering the classical configurations of
the field-theoretical construction, it is quite difficult to examine
the quantum effect.
Another example is the domain wall that is induced by the permutation
between branes. 
As far as we are considering only the classical brane configurations,
the permutation between branes seems to generate another
degenerated vacuum\cite{incidental}.
In this section, however, we show that such permutations become trivial
if the quantum effect of the brane dynamics is properly
included.\footnote{From the viewpoint of the effective four-dimensional
action, the 
triviality of the permutation 
is originated from the gauge fixing, which is unclear in its
classical counterpart of the brane dynamics. See appendix for more detail.}
Then we discuss about brane Q-balls.
In the usual four-dimensional theory, one can construct Q-balls if 
\begin{itemize}
\item There is a flat potential that is lifted by a small perturbation.
\item There is an effective $U(1)$ global symmetry.
\end{itemize}
Our model satisfies the above criteria. 
The brane Q-balls are conceptually different from the conventional
Q-balls\cite{BraneQball}.
We show how one can distinguish the conventional Q-balls from the brane
Q-balls. 

We show that cosmological domain walls, axionic strings and Q-balls can be
produced by the usual brane dynamics after brane inflation.
We stress here that it is quite natural to expect cosmological formation
of these defects in the conventional scenarios of the evolution of the
Universe.

\subsection{Brane Dynamics}
The analysis of the four-dimensional effective action is discussed in
appendix A. 
We pay attention to the discussions in this section so that 
they become parallel to the analysis in appendix A.  

Here we consider brane configurations of the $3+1$
dimensional $N=2$ supersymmetric Yang-Mills (SYM), and the brane
counterpart of the cosmological defects.
The starting point of our discussion is the brane configuration of type
IIA string theory consisting of solitonic (NS) five-branes and $N_c$
D4-branes.
Following ref.\cite{MQCD-review}, we consider the object;
\begin{eqnarray}
NS5 && (x^0, x^1, x^2, x^3, x^4, x^5)\nonumber\\
D4 &&  (x^0, x^1, x^2, x^3, x^6),
\end{eqnarray}
which is schematically sketched in fig.\ref{MQCD_NS5-D4}.
Here the gauge coupling of the $3+1$ dimensional gauge theory is given by;
\begin{equation}
\frac{1}{g^2}=\frac{L}{g_s l_s}.
\end{equation}
If supersymmetry is not broken, there are no forces between D4-branes,
and the field that parameterizes the brane distance represents the flat
direction of the effective four-dimensional Lagrangian.
However, in more realistic situations, supersymmetry must be broken.
Here we consider a simple assumption that a tiny breaking of
supersymmetry  
induces weak repulsive force between the D4-branes.
In this case, the thermal effect can stabilize the potential
between branes, which induce the restoration of the corresponding 
symmetry in the four-dimensional effective
action\cite{incidental, thermal-brane}.
After symmetry breaking, D4-branes are placed on a circle, as is
depicted in fig.\ref{basic-brane}.\footnote{Here we have assumed that the
higher dimensional terms stabilize the potential at a distance, as
we have considered in the discussion about 
four-dimensional theory in appendix A.}
The elements ($a_i$'s) in the effective theory correspond to the
position $(x^4, x^5)$ of the endpoints of the D4-branes on NS5.

Let us first consider a naive permutation between the two D4-branes.
%Experts of MQCD might think that such discussions are useless,
%however 
We stress here that it is important to understand why the naive
permutations in the classical configuration(see 
fig.\ref{Permutation_Wall}) {\bf do not} induce
any physical domain walls.
It should be noted that it is rather hard to understand the reason
from the classical configuration.
As we have discussed for the effective action,
the quantum effect is described by the deformation of the moduli
space.
The vacuum of the MQCD coulomb branch\cite{MQCD-review} is represented
by the  
curve, which takes the same form as eq.(\ref{curve-SYM}).
From the viewpoint of the brane dynamics, the curve represents a
M5-brane.
In this case, one can see that the above-mentioned permutation in the
classical configuration becomes a trivial symmetry of the quantum
configuration, which works on the two parts of the identical M5-brane.
This result is a triumph of MQCD, as it cannot be obtained
from the classical argument of the brane dynamics.
Moreover, as is already discussed in ref.\cite{Witten_Wall}, defects are
the important probes for the examination of the consistency between
SQCD and MQCD.
In our case, since the results obtained from the quantum brane dynamics are
consistent with SQCD for the axionic strings and the domain walls, one
can understand that 
our results present alternative proofs for the consistency
between SQCD and MQCD in a rather exotic situation.  

Although the permutations do not induce
any physical domain wall,
the axionic string and $Z_{N_c}$ domain walls are still
the physical objects of the quantum brane dynamics.\footnote{Unlike
the usual BPS domain walls in $N=1$ SQCD, these
defects are unstable in the supersymmetric limit. 
On the other hand, the usual BPS domain walls cannot appear in the
classical configuration.}
We have included a small soft mass that destabilizes the coulomb branch,
which is so small that it does not ruin the equations for the curve.
The classical $U(1)_R$ symmetry is broken by curving the left and the right
fivebranes to 
\begin{eqnarray}
s_L&=&- N_c R_{10} \log v\nonumber\\
s_R&=&  N_c R_{10} \log v,
\end{eqnarray}
where $s= x^6 + i x^{10}$, and $R_{10}$ is the radius of the
compactified $x^{10}$.
Because of the fact that $Im s = x^{10}$ lives on a circle of radius
$R_{10}$, there is the residual discrete symmetry $Z_{N_c}$.
The domain walls are produced by the 
spontaneous breaking of $Z_{N_c}$. 
Similar to the usual cases, our axionic strings are formed at the 
scale where the spontaneous breaking of the $U(1)_R$ symmetry is induced
by $<\phi>\ne 0$.
Then a dynamical potential is generated at lower energy scale, which
breaks $U(1)_R$ to the discrete symmetry.
At this time, the axionic string becomes a junction (or boundary) of
the $Z_{N_c}$ domain walls.\footnote{Cosmological considerations of
the stability of the usual BPS domain walls are
 already discussed in ref.\cite{Matsuda-weak-wall}.}

Here we should note how one can circumvent the criteria
given in ref.\cite{Brane-defects, angled-defect, D-brane-strings, Halyo,
BDKP-FI}, in which the formation of cosmological domain walls is
discussed to be negligible. 
In our case, although it is not impossible to think that the domain walls are
the daughter branes being extended between vacuum branes,
cosmological consideration suggests that they
are formed by the continuous deformations of the vacuum branes,
therefore they are not produced by tachyon condensation.
The actual construction of the $Z_{N_c}$
domain walls is straightforward. 
Since the supersymmetry breaking is weak in this case, 
thermal effect can stabilize the potential between branes.
At the beginning when the temperature is higher than the dynamical
scale, the branes are placed on top of each other. 
The corresponding gauge symmetry is restored in the effective action.
Then the branes start to fall apart at low temperature, with spatial
fluctuations of their positions as is anticipated by the conventional
Kibble mechanism.
The domain walls are described by the fivebranes that interpolate
between the two adjacent vacua.
Let us consider a domain wall, which looks like one vacuum of the theory
for $z\rightarrow -\infty$ and looks like another vacuum for
$z\rightarrow \infty$.
Here we denote these two vacua by the index ``A'' and ``B'', and
$z$ is one of the three spatial coordinates.
Then the domain wall is described by the fivebrane that has two
adjacent vacua on its boundaries, morphing one to the other along the
z direction.
If the vacuum states were described by the fivebranes of the form $R^4\times
\Sigma$, where $\Sigma$ was a Riemann surface embedded in the extra
dimensions $Y \equiv R^5 \times S^1$, the domain walls are described
by the fivebranes of the form $R^3 \times D$, where $R^3$ is the
four-dimensional spacetime without $z$, and $D$ is a three-surface in
the seven manifold $Y' \equiv R_z \times Y$, where $R_z$ is the copy of
the spatial $z$ direction.
Near $z\rightarrow -\infty$, $D$ should look like $R_z \times \Sigma_A$.
On the other side, near $z\rightarrow \infty$, $D$ should look as $R_z
\times \Sigma_B$.
The defect is described by the continuous 
deformation of the existing M-theory fivebrane, morphing from one side
to the other along the $z$ direction.
The cosmological formation of such defects is already discussed in
ref.\cite{incidental}, by using the field-theoretical construction.

{\bf As we have stated above, in addition to the conventional
brane defects that are formed by brane creation,
one should consider another kind of brane defects that are formed by the
continuous deformations of the branes if one wants to consider generic
cases of cosmological scenarios.  
Since the two kinds of brane defects can be produced 
by the same process, one must deal with the mixture of these defects
in the actual analysis of the brane Universe.}

Finally, we consider the brane Q-balls in the Hanany-Witten type brane
dynamics.
Conventional Q-balls in the four-dimensional theories 
are already discussed by many authors\cite{Q-ball-sugra}.
A brane-counterpart of the Q-ball is discussed in \cite{BraneQball}
for brane anti-brane pair.
Here we consider the brane Q-balls in the Hanany-Witten type brane
dynamics.
Our discussions are parallel to the analysis of the effective action in
appendix A.
Unlike the above discussions for axionic strings and domain walls,
since the brane Q-balls are constructed deep inside the classical region,
the configuration of the brane Q-ball is not affected by the quantum
effect.
However, as is discussed in \cite{BraneQball}, there is a {\bf crucial
difference between Q-balls in the effective action and brane
Q-balls}.
In the case of brane anti-brane pair, the decay mode of the brane Q-ball
is dominated by the radiation into the bulk when 
the charge of the brane Q-ball exceeds a critical value.
In the followings, we derive the critical charge in the Hanany-Witten
type brane dynamics. 
We assume that the supersymmetry breaking is
dominated by the conventional supergravity mediation at large $<\phi>$.
In the cases of the gravity mediation, the potential is schematically
given by\footnote{Here we assume that $M_1 > \Lambda$, where $\Lambda$
denotes the scale where the dynamical effect becomes dominant.
We also assume that $\phi_Q > \sqrt{2} M_1^2/m_{3/2}$,
so that the gravitational mediation becomes effective.
$M_1$ is defined in appendix A.}
\begin{equation}
\label{pot_sug}
V(\phi) = m_{3/2}|\phi|^2 \left(1+K \log \frac{|\phi|^2}{M_*^2}\right).
\end{equation}
Then the Q-balls have the properties,
\begin{eqnarray}
\label{braneQ-sugra}
R_Q \simeq |K|^{-1/2}m_{3/2}^{-1} &,& \omega \simeq m_{3/2} \nonumber\\
|\phi_Q| \simeq |K|^{3/4}m_{3/2}Q^{1/2} &,& E_Q \simeq m_{3/2} Q,
\end{eqnarray}
When the branes are rotating, the acceleration of the rotating branes
in the bulk ( which is denoted by $a_b$) generates the radiation
into the bulk of the 
form\cite{branonium},
\begin{equation}
\label{bulk-rad}
\left|\frac{d E_Q}{dt}\right|\simeq \frac{1}{8\pi}
(\kappa_4 T_{D4} V_{D4})^2 a_b^2, 
\end{equation}
where $\kappa_4$ and $V_{D4}$ are the four-dimensional gravitational
coupling and the spatial volume of the D4-brane, which becomes $V_{D4}
\simeq \frac{4\pi}{3}R_Q^3 \times L$ in our model.
Here $L$ is the length of the D4-brane in the compactified direction.
$T_{D4}$ is the tension of the D4-brane, which is given by the usual
formula, $T_{D4} \simeq M_*^5$.
Here we consider the acceleration of the rotating brane that is given by
$a_b \simeq (|\phi_Q|/M_*^2) \omega^2$.
On the other hand, the conventional decay mode has been
studied in ref.\cite{Q-ball-evapolation}, which is given by;
\begin{equation}
\label{norm}
\left|\frac{dQ}{dt}\right| \le \frac{\omega^3 4\pi R_Q^2}{192\pi^2}.
\end{equation}
From eq.(\ref{braneQ-sugra}) and eq.(\ref{norm}), one can obtain;
\begin{equation}
\label{normal-rad}
\left|\frac{d E_Q}{dt}\right| \le \frac{m_{3/2} \omega^3 4\pi R_Q^2}{192\pi^2}.
\end{equation}
From eq.(\ref{bulk-rad}) and (\ref{normal-rad}), we can derive the
critical charge
\begin{equation}
Q_c \simeq 10^{-2}\times \frac{K^{1/2}m_{3/2}^2}{\kappa_4^2 L^2 M_*^6}.
\end{equation}
The above result suggests that the decay of the brane Q-ball is {\bf always}
dominated by the efficient radiation into the bulk. 
The existence of the critical charge is crucial in the analysis of 
realistic cosmological models.
If the actual gauge symmetry of the Universe is due to the brane dynamics of
the Hanany-Witten type, Q-balls should have their brane
counterpart, which {\bf can be distinguished from the conventional
Q-balls even if the four-dimensional effective action looks the same}.

Since we are considering large Q-balls, we should mention the relation
between the scales of the compactified space and large $\phi_Q$.
In the above example, the gravitino mass is given by $m_{3/2}\simeq 
\Lambda_{susy}^2/M_p$, where $\Lambda_{susy}$ denotes the scale of the
supersymmetry breaking.
Although the charge of the Q-ball becomes quite large in conventional
cosmological scenarios, the distance between rotating branes, which is
denoted by the scalar field $\phi_Q$, do not exceed the compactified
radius.
In this case, the suppression factor of $m_{3/2}$ in
eq.(\ref{braneQ-sugra}) is crucial.
Therefore, because of the factor $m_{3/2}$ in eq.(\ref{braneQ-sugra}),
we can safely 
embed the configuration of brane Q-balls in the above brane model.

\section{Conclusions and Discussions}
In this paper, we have considered the formation of D-term strings,
axionic strings, domain walls and Q-balls in the Hanany-Witten type brane
dynamics. 
Here we summarize our conclusions.
\begin{itemize}
\item For the D-term string, we have considered D2-brane that is stretched
      between the splitting D4-branes.
      Contrary to the previous arguments, the production of the extended
      D2-branes is not suppressed. 
      Our arguments are general, because the brane collision with a
      huge kinetic energy will inevitably induce chaotic process of the 
      production/annihilation and the recombination of the branes, which
      makes 
      it possible to produce many kinds of extended daughter branes.
      Further discussions of this topic is given in
      ref.\cite{matsuda_angleddefect, matsuda_future}.
\item We have considered the production of axionic strings and domain walls.
      In our case, the defects are not produced by the tachyon
      condensation but are formed by the spatial deformations of the
      mother branes. 
      The parameter of the deformation is the position of the branes in
      the compactified space.
      These defects are first constructed in the classical brane
      configuration, and then lifted to MQCD.
      We have shown that quantum effect is crucial for axionic strings
      and domain walls.
      On the other hand, since the defects are  the non-trivial
      excitations of the system, they are also important in examining the
      consistency between SQCD and MQCD\cite{Witten_Wall}.
\item We have discussed brane Q-balls in the Hanany-Witten brane
      dynamics. We have found that there is a distinguishable difference
      between brane Q-balls and conventional Q-balls.
\end{itemize}

In our future work\cite{matsuda_future}, it is shown that 
monopoles and domain walls can be produced by daughter brane
creations, which are extended between mother branes.
For the cosmic strings, in our next paper\cite{matsuda_angleddefect}, the
consideration of the extended daughter 
brane is used to solve the long-standing problem of the
$\theta$-dependence. 
{\bf In either case, the production of the extended brane is crucial for
the cosmological defect formation}.
We show in ref.\cite{matsuda_future} that the cosmological process that
is required for the creation of the monopoles and the domain walls
is the same as the one required for the formation of incidental brane
defects.
{\bf Then, the actual cosmological relics after brane inflation are the
mixture of the two kinds.}
The cosmological evolution of such brane defects is quite interesting
and deserves further discussions.

\section{Acknowledgment}
We wish to thank K.Shima for encouragement, and our colleagues in
Tokyo University for their kind hospitality.

\appendix
\section{Four-dimensional N=2 SYM and coulomb branch}

\begin{figure}[ht]
 \begin{center}
\begin{picture}(130,200)(0,0)
\resizebox{5cm}{!}{\includegraphics{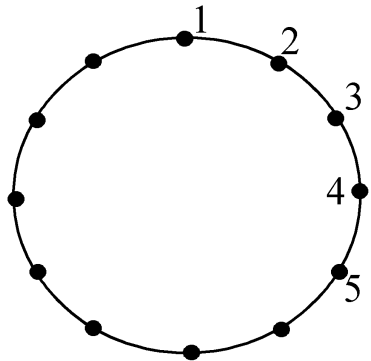}} 
\end{picture}
\caption{Each dot correspond to the position of the branes in the
  codimension 2 space.
In the effective four-dimensional Lagrangian, the dots represent $a_i$'s.}  
\label{basic-brane}
 \end{center}
\end{figure}

\begin{figure}[ht]
 \begin{center}
\begin{picture}(300,200)(0,0)
\resizebox{10cm}{!}{\includegraphics{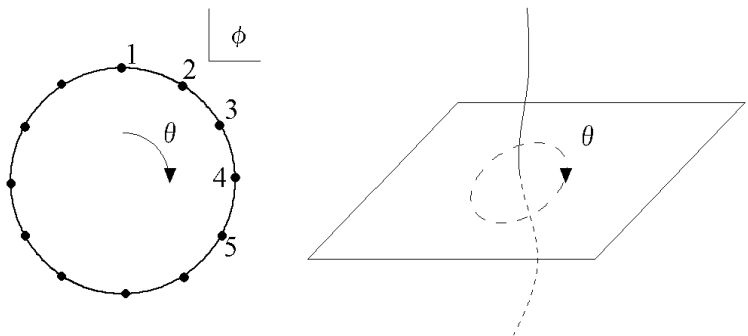}} 
\end{picture}
\caption{The $U(1)_R$ rotation of the adjoint scalar field is shown in
  the left picture. 
If the windings are formed in the three-dimensional space, they 
form strings as is depicted in the right picture.} 
\label{Axionstring}
 \end{center}
\end{figure}

$N=2$ supersymmetric gauge theory with the gauge group $SU(N_c)$ is
written by $N=1$ supersymmetric vectormultiplet and a chiral superfield 
in the adjoint representation\cite{MQCD-review}.
\begin{equation}
\label{4D-N2-SYM}
{\cal L}_{vec} = Im \, Tr \left[
\tau \left(\int d^4 \theta \Phi ^{\dag} e^{-2V}\Phi
+ \int d^2 \theta W_\alpha W^\alpha 
\right)
\right]
\end{equation}
where the trace runs over the gauge group, and 
$\tau=\frac{\theta}{2\pi}+\frac{i}{g^2}$ is the complex
coupling.
In components, the bosonic part of the superfield $\Phi$ includes the
potential
\begin{equation}
V \sim Tr[\phi^{\dag}, \phi]^2,
\end{equation}
which has flat directions.
The Lagrangian (\ref{4D-N2-SYM}) is invariant under the $U(1)_R$
symmetry $\Phi\rightarrow e^{2 i \alpha_R}\Phi(e^{-i \alpha_R}\theta)$
that is a consequence of the classical conformal invariance.
Here we assume that the Fayet-Iliopoulos D-term is zero.
The vacuum expectation value of the complex scalar field $\phi$, which is the
lowest component of the adjoint superfield $\Phi$, can always be
rotated by a gauge  
transformation, to lie in the Cartan subalgebra of $SU(N_c)$.
Namely, one can always rotate the vev to be in the form
\begin{eqnarray}
\label{VEV-adjoint}
<\phi> & = & \sum_{i=1} a_i H_i\nonumber\\
       & = & diag[a_1, a_2, ..., a_{N_c}].
\end{eqnarray}
Up to gauge transformation, the D-flatness condition
$[\phi^{\dag},\phi]=0$ is satisfied when
\begin{equation}
 \sum_{i=1} a_i = 0. 
\end{equation}
The elements of the $SU(N_c)$, which act non-trivially on the Cartan
subalgebra, are the elements of the Weyl group, isomorphic to the
permutation group $S_{N_c}$.
We assume that there is a mechanism that
destabilizes the origin of the potential, and the elements are placed
along a circle.\footnote{For example, one can assume that the supersymmetry is
broken 
in a hidden sector, and the breaking of supersymmetry is induced by the
soft term 
for the field $\Phi$, 
\begin{equation}
V_{soft}=-\frac{1}{2} m_{soft}^2 Tr |\phi|^2
\end{equation}
which destabilizes the coulomb branch.
Because of the destabilized potential, $\phi$ develops large vacuum
expectation value.}
We also assume that the stabilization of the potential at large
$\phi$ is due to the
nonrenormalizable terms.
In this case, the vacuum expectation value of $\phi$, which satisfies the
traceless condition $\sum_i  a_i =0$, is given by 
\begin{equation}
a_k = |a| e^{\frac{2\pi i}{N_c} k}.
\end{equation}
We are considering the case where the soft breaking of supersymmetry
induces spontaneous breaking of the $SU(N_c)$ symmetry in the classical 
vacuum.
The $a_i$'s in (\ref{VEV-adjoint}) are distributed on a circle as is
depicted in fig.\ref{basic-brane}. 
The global $U(1)_R$ symmetry rotates the field as
$\Phi\rightarrow e^{2 i \alpha_R}\Phi(e^{-i \alpha_R}\theta)$,
which is the symmetry that induces the axionic string depicted in
fig.\ref{Axionstring}.  
The $U(1)_R$ symmetry is broken by
the quantum effect to the discrete symmetry, $U(1)_R \rightarrow
Z_{4N_c}$.
Since the superfield $\Phi$ is charged 2 under the $U(1)_R$ symmetry, the
effective symmetry of the field $\phi$ is $Z_{2N_c}$.
The spontaneous breaking of the discrete symmetry is the origin of the
 domain walls.
However in the present model, the $N_c$ domain walls are produced by the
spontaneous breaking of the discrete $Z_{N_c}$ symmetry, not by the
discrete symmetry $Z_{2N_c}$. 
To see the mechanism of the domain wall formation, we briefly
review the structure of the coulomb branch.
We are assuming that the soft supersymmetry breaking is so weak that
the usual analysis on $N=2$ coulomb branch is still a good approximation.
The curve\cite{MQCD-review} is described by,
\begin{equation}
\label{curve-SYM}
t^2+P_n(v) t +1=0.
\end{equation}
Here $P_n(v)$ is the polynomial of the form
\begin{equation}
\label{finite-v}
P_n(v) = v^n + u_2 v^{n-2} + ...+u_n,
\end{equation}
where $u_i$'s are the order parameters of the theory, which are usually
given by the formula $u_i=Tr\Phi^i$.
For large $v$, the above equations behave like
\begin{equation}
\label{infinite-v}
t_{\pm} \simeq v^{\pm N_c},
\end{equation}
where $t_\pm$ denotes the two roots of $t$.
The $U(1)_R$ symmetry rotates $v$ that has the $U(1)_R$
charge of 2.
However, the remaining $Z_{2N_c}$ symmetry of $v$ is not the symmetry of
(\ref{infinite-v}).
Considering the discrete rotation of the form
\begin{eqnarray}
v' &\rightarrow& v e^{\frac{2\pi i}{2N_c}k}\nonumber\\
&&k=1,2,.., 2N_c,
\end{eqnarray}
the $v^{'\pm N_c}/v^{\pm N_c} = -1 $ element does not keep
(\ref{infinite-v}) invariant. 
Thus the symmetry of (\ref{infinite-v})
becomes $Z_{N_c}$, which is spontaneously broken in (\ref{finite-v}).
The spontaneously broken discrete symmetry $Z_{N_c}$ induces $N_c$
domain walls.

Here we  consider a rather trivial question.
Why the domain walls are not induced by the permutation between $a_i$'s ?
Although the reason is easily understood if one considers the gauge
fixing, we will show another answer, without mentioning the
gauge fixing.
One may think that our arguments are circumbendibus, however
they are useful for our later discussions about the brane
dynamics. 
In the quantum space of eq.(\ref{finite-v}), the permutation corresponds
to the permutation between the $N_c$ roots $a_i$.
Of course, the permutation is the exact symmetry of the curve
(\ref{finite-v}), which is never broken.
Thus in the quantum moduli space, the permutation cannot induce domain
walls, because the permutation 
symmetry is trivial.

Finally, we consider the Q-balls in this four-dimensional effective
Lagrangian.
Unlike the setups for axionic string and domain walls, the
origin of the flat direction is needed to be the global minimum.
We assume that the potential is slightly lifted by a weak
supersymmetry breaking, or a small $\mu$-term such as $\mu Tr \Phi^2$,
so that the origin of $\phi$ becomes the global minimum. 
There remains an effective $U(1)_R$ global symmetry, which is later
broken to $Z_{4N_c}$ by the quantum effect.
In such situations, the Q-balls can be produced\cite{Q-ball-sugra} if 
the potential that breaks the global $U(1)_R$ symmetry is
shallow.\footnote{The situation is similar to the conventional QCD
axion.}
The following potential will be generated for the adjoint
field $\phi$, 
\begin{equation}
V_{adj} = M_1^4 \log \left(1+\frac{|\phi|^2}{M_1^2}\right)
+ m_{3/2}^2 |\phi|^2 
\left(1+ K \log 
\frac{|\phi|^2}{M_*^2}
\right),
\end{equation}
where $m_{3/2}$ is the gravitino mass,
and $K$ is determined by the one-loop corrections.
Here $M_1$ is determined by the mechanism of the supersymmetry breaking
(which is not fixed in our discussion),
and $M_*$ is the renormalization scale.
Using the conventional notations $\phi = \phi_Q e^{i\omega t}$,
where $\phi_Q$ denotes the vacuum expectation value of the field 
$\phi$ in the core of the Q-ball, one can find for the large
Q-ball\cite{Q-ball-sugra}, 
\begin{eqnarray}
R_Q \simeq |K|^{-1/2}m_{3/2}^{-1} &,& \omega \simeq m_{3/2} \nonumber\\
\phi_Q \simeq |K|^{3/4}m_{3/2}Q^{1/2} &,& E_Q \simeq m_{3/2} Q,
\end{eqnarray}
for $\phi_Q \,>\, \sqrt{2}M_1^2/m_{3/2}$, and
\begin{eqnarray}
R_Q \simeq M_1^{-1}Q^{1/4} &,& \omega \simeq M_1 Q^{-1/4}\nonumber\\
\phi_Q \simeq M_1 Q^{1/4} &,& E_Q \simeq M_1 Q^{3/4},
\end{eqnarray}
for $\phi_Q \, <\, \sqrt{2}M_1^2/m_{3/2}$.
Here $Q$, $R_Q$ and $E_Q$ are the charge, the radius
and the energy of the Q-ball. 
We consider the brane counterpart of the above-mentioned Q-balls
in the latter half of section 3.

\end{document}